\documentclass{iopart}

\usepackage{ifthen}
\usepackage{ifpdf}

\usepackage{latexsym}
\usepackage{amssymb}
\usepackage{bm}

\ifpdf
\usepackage{graphicx}
\usepackage{epstopdf}
\else
\usepackage{graphicx}
\usepackage{epsfig}
\fi

\newcommand{\eexp}{\mbox{e}^}

\newcommand{\mass}{\mathsf{m}}

\newcommand{\tbox}[1]{\mbox{\tiny #1}}
 
\newcommand{\amatrix}[1]{\matrix{#1}} 

\newcommand{\be}[1]{\begin{eqnarray}\ifthenelse{#1=-1}{\nonumber}{\ifthenelse{#1=0}{}{\label{e#1}}}}
\newcommand{\ee}{\end{eqnarray}} 

\newcommand{\hide}[1]{}

\newcommand{\mpg}[2][1.0\hsize]{\begin{minipage}[b]{#1}{#2}\end{minipage}}

\newcommand{\putgraph}[2][width=0.30\hsize]{\includegraphics[#1]{#2}}

\begin{document} 

\title{The mesoscopic conductance of ballistic rings} 

\author{
Yoav Etzioni, 
Swarnali Bandopadhyay\footnote{Present address: 
Max Planck Institute for the Physics of Complex Systems,
N{\"o}thnitzer Str. 38, 01187 Dresden, Germany}
and Doron Cohen}

\address{
Department of Physics, Ben-Gurion University, Beer-Sheva 84005, Israel
}

\begin{abstract}
The calculation of the conductance 
of ballistic rings requires a theory 
that goes well beyond the Kubo-Drude formula.  
Assuming ``mesoscopic" circumstance of  
very weak environmental relaxation, 
the conductance is much smaller compared 
with the naive expectation. Namely,   
the electro-motive-force induces an energy 
absorption with a rate that depends crucially 
on the possibility to make connected sequences 
of transitions. Thus  the calculation 
of the mesoscopic conductance is similar 
to solving a percolation problem. 
The ``percolation" is in energy space 
rather than in real space. 
Non-universal structures and sparsity 
of the perturbation matrix cannot be ignored. 
The latter are implied by lack of 
quantum-chaos ergodicity in ring shaped 
ballistic devices.
\end{abstract}

\section{Introduction}

Closed mesoscopic rings are of great interest~[1-13].
For such devices the relation between 
the conductance and the internal dynamics 
is understood much less~\cite{kbf,kbr} 
than for open systems. 
First measurements of the conductance of closed 
mesoscopic rings have been reported more than 
a decade ago~\cite{orsay}.  
In a typical experiment a collection 
of mesoscopic rings is driven by 
a time dependent magnetic flux $\Phi(t)$ which creates 
an electro-motive-force (EMF) ${-\dot{\Phi}}$ in each
ring. Assuming that Ohm's law applies, the induced current 
is ${\mathcal{I}=-G\dot{\Phi} }$ and consequently the rate 
of energy absorption is given by Joule's law as
\be{1}
\dot{\mathcal{W}} \ \ = \ \  
\mbox{Rate of energy absorption} 
\ \ = \ \ G\,\dot{\Phi}^2 
\ee
where $G$ is called the conductance 
\footnote{The terminology of this paper, and in particular 
our notion of ``conductance" are the same as in the 
theoretical review \cite{kamenev} 
and in the experimental work \cite{orsay}.}.
For diffusive rings the Kubo formalism 
leads to the Drude formula for~$G$.  
A major challenge in past studies 
was to calculate the weak localization 
corrections to the Drude result, 
taking into account the level statistics  
and the type of occupation \cite{kamenev}. 
It should be clear that these corrections 
do not challenge the leading order Kubo-Drude result. \\

{\bf Theoretical challenge:} 
It is just natural to ask~\cite{kbf,kbr} 
what happens to the Drude result if the disorder 
becomes weak (ballistic case) or strong 
(Anderson localization case). 
In both cases the individual eigenfunctions  
become non ergodic: a typical eigenfunction 
do not fill the whole accessible phase space. 
In the ballistic case a typical eigenfunction  
is not ergodic over the open modes in momentum space,  
while in the strong localization case 
it is not ergodic over the ring in real space.  \\

{\bf Major observation:}
Lack of quantum ergodicity implies that 
the perturbation matrix 
is very structured and/or sparse. 
Consequently the calculation of~$G$ 
requires a non-trivial extension  
of linear response theory (LRT). 
Such extension has been proposed in Ref.\cite{kbr}  
and later termed ``semi linear response theory" (SLRT) \cite{slr1,slr}.    
As in the standard derivation of the Kubo formula, 
also within the framework of SLRT, 
the leading mechanism for absorption 
is assumed to be Fermi-golden-rule (FGR) transitions. 
These are proportional to the squared matrix 
elements $|\mathcal{I}_{nm}|^2$ of the current operator. 
Still, the theory of \cite{kbr} 
does not lead to the Kubo formula. 
This is because the rate of absorption depends 
crucially on the possibility to make {\em connected} 
sequences of transitions, and it is greatly reduced 
by the presence of bottlenecks. 
It is implied that both the structure 
of the $|\mathcal{I}_{nm}|^2$ band profile 
and its sparsity play a major role in the calculation of $G$. \\

{\bf SLRT and beyond:}
Within SLRT it is assumed that the transitions 
between levels are given by Fermi-golden-rule, 
but a resistor network analogy \cite{miller} 
is used in order to calculate the overall absorption. 
The calculation of the energy absorption 
in Eq.(\ref{e1}) is somewhat similar to solving 
a percolation problem. The ``percolation" is in 
energy space rather than in real space. 
A recent study \cite{silva} suggests a way to go 
beyond the FGR approximation. If the results 
of Ref~\cite{silva} could be extended beyond the 
diffusive regime \cite{silvaPC} it would be possible 
to extend SLRT into the non-linear regime.  \\

{\bf Scope:}
In a follow up work \cite{kbv} its is demonstrated  
that for a very strong disorder SLRT leads naturally 
to the resistor network ``hopping" picture \cite{ambeg,pollak}, 
from which Mott's variable-range-hopping 
approximation \cite{mott1,mott2} can be derived.  
In the present work we apply SLRT to the other extreme case, 
of having a ballistic ring.  
It should be appreciated that the generalized resistor 
network picture of SLRT provides a firm unified framework 
for the calculation of the mesoscopic conductance:
The same recipe is used in both extreme cases 
without the need to rely on an ad-hock phenomenology.  \\

{\bf Ballistic rings:}
In the case of a diffusive rings the mean free path~$\ell$ 
is assumed to be much less than the perimeter~$L$ of the ring.
In the case if ballistic ring we have the opposite 
situation $\ell \gg L$. One way to model a ballistic 
ring is to consider very weak disorder (Fig.~1a). 
Another possibility is to consider a clean ring with 
a single scatterer (Fig.~1b) or a small deformation (Fig.~1c).  
In the latter cases it is natural to characterize the 
scattering region by its total transmission~${g_T \sim 1}$. 
Consequently the mean free path is  
\be{2}
\ell \ \ \approx \ \ \frac{1}{1-g_T} L
\ee
The assumption of ballistic motion implies 
that ``quantum chaos" considerations 
are important. Consequently one should go well beyond 
the conventional random wave picture of Mott~\cite{mott1,mott2}, 
and beyond the standard Random Matrix Theory analysis.     
In fact we are going to explain that in this limit 
the eigenfunctions are non-ergodic, and hence 
the perturbation matrix $|\mathcal{I}_{nm}|^2$
is structured and sparse. \\

\newpage

{\bf Outline:} The purpose of this paper is 
to discuss the modeling of ballistic rings; 
to clarify the procedure which is involved 
in the calculation of the conductance;   
and to analyze the simplest example.
In sections~2 we distinguish between  
\begin{itemize}
\item Disordered rings (e.g. Anderson model)
\item Chaotic rings (e.g. billiard systems)  
\item Network models (also known as ``graphs")
\end{itemize}
These are illustrated in Fig.~1.
In particular we motivate the analysis of a simple 
prototype network model for a multimode ring. 
This model has all the essential ingredients 
to demonstrate the major theme of this paper.

In sections~3-8 we elaborate on the procedure 
which is used for the calculation of the conductance. 
This is not merely a technical issue, 
since new concepts \cite{kbr,slr} are involved. 
We make a distinction between:
\begin{itemize}
\item The Landauer result for the open device 
\item The classical Drude result for a closed ring
\item The (quantum) spectroscopic conductance of a ring
\item The (quantum) mesoscopic conductance of a ring
\end{itemize}
Here  "classical" as opposed to ``quantum"   
should be understood in the sense of Boltzmann picture. 
In the ``classical" case the interference within  
the arm of the ring is ignored, while both the Fermi statistics 
and the single scattering events are treated properly.   
For the prototype model we get the following 
simple results: Given that the device has $\mathcal{M}$ 
open modes and its total transmission is $g_T<1$, 
the Landauer conductance is~\cite{stone} 
\be{3}
G_{\tbox{Landauer}} = \frac{e^2}{2\pi\hbar} \mathcal{M} \ g_T
\ee
The Drude result for the closed ring is 
\be{4}
G_{\tbox{Drude}} = \frac{e^2}{2\pi\hbar} \mathcal{M} \ \frac{g_T}{1-g_T}
\ee
and the associated quantum results are
\be{5}
G_{\tbox{spec}} &\approx& \frac{e^2}{2\pi\hbar} \mathcal{M}
\times \mbox{minimum}\left[\frac{g_T}{1-g_T}, \mathcal{M}\right]
\\ \label{e6}
G_{\tbox{meso}} &=& \frac{e^2}{2\pi\hbar} \ 2\mathcal{M}^2 \ \mathsf{g}_{\tbox{meso}}
\ee
The calculation of $\mathsf{g}_{\tbox{meso}}$ 
involves a complicated {\em coarse graining} 
procedure that we discuss in section~8. 
The spectroscopic result $G_{\tbox{spec}}$ 
describes via Eq.(\ref{e1}) both the initial 
(transient) rate and also the long time 
(steady state) rate of energy absorption,  
provided the environment provides a strong 
relaxation mechanism. 
The mesoscopic result describes the 
(slower) long time rate of energy absorption 
if the environmentally induced relaxation is weak. 
See \cite{kbr} for an extended quantitative discussion.

The outcome for $G_{\tbox{meso}}$  may differ 
by orders of magnitude from the conventional Kubo-Drude result.
The calculation procedure implies that 
${G_{\tbox{meso}} < G_{\tbox{spec}} \le  G_{\tbox{Drude}}}$.  
In the last part of this paper (sections~9-12) we demonstrate  
this point via the analysis of the prototype model. 
Our numerical results, whose preliminary version 
were reported in~\cite{bls}, suggest that 
typically ${ G_{\tbox{meso}} <  G_{\tbox{Landauer}} }$. 
The results of the calculation are contrasted 
with those of the conventional Kubo approach, 
and their {\em robustness} is discussed.

\section{Modeling}

A simple model for a ballistic ring can be either 
of the``disordered type"  or of the ``chaotic type". 
Let us visualize the disordered potential as 
arising from a set of scatterers which are 
distributed all over the ring (Fig~1a). 
Depending on the scattering 
cross section of the individual scatterers we can have 
$\ell \ll L$ for strong disorder or $\ell \gg L$ 
for weak disorder, where $\ell$ is the mean free path 
for velocity randomization, and $L$ is the perimeter of 
the ring.

Another possibility is not to make all the scatterers    
smaller, but rather to dilute them. 
Eventually we may have a chaotic ring where 
the scattering is induced by a single scatterer (Fig~1b).  
For example the scatterer can be a 
disc or a semi-disc as in Fig~2. 
These variations of Sinai billiard   
(billiard with convex wall elements)
are known to be chaotic.
It is important to remember that ``chaos" 
means that complicated ergodic 
classical dynamic is generated by a simple 
Hamiltonian ({\em no disorder!}).

In our view chaotic rings are more 
interesting for various reasons. 
Ballistic devices are state-of-art 
in mesoscopic experiments. 
For example it is quite common 
to fabricate Aharonov-Bohm devices. 
In such devices it is possible to induce local 
deformation of the potential by means of 
a gate voltage. Hence one has a full control over 
the amount of scattering. Also from theoretical 
point of view it is nice to have a well defined 
scattering region: This allows to use the 
powerful $S$-matrix point of view that has been 
initiated by Landauer. In particular we can ask 
what is the conductance of a device depending on 
whether it is integrated in an open geometry 
as in Fig.~1e or in a closed geometry as in Fig.~1d.   
We believe that ``chaos" and ``disorder" lead 
to similar physics in the present context, but this
claim goes beyond the scope of the present paper.

A multimode ring can be visualized 
as a waveguide of length $L$ and width $W$.  
In such case the number of open modes 
is $\mathcal{M} \propto (k_{\tbox{F}} W)^{d{-}1}$ 
where $d=2,3$ is the dimensionality.
We label the modes as 
\be{0}
a \ \ = \ \  \mbox{mode index}  \ \ = \ \ 1,2,...,\mathcal{M}
\ee
The scattering arise due to some bump or 
some deformation of the boundary,  
and can be described by 
an $2\mathcal{M} \times 2\mathcal{M}$ 
scattering matrix~$\bm{S}$. For the semi-disc 
model analytical complicated expressions 
are available \cite{blumel,holger}.  
The ``classical" transitions probability matrix $\bm{g}$ 
is obtained by squaring the absolute values 
of the $\bm{S}$ matrix elements.

Disregarding the closed channels,  
the ballistic ring is described  
as a set   of $\mathcal{M}$ open modes, 
and a small scattering region 
that is characterized by its total transmission~$g_T$. 
Optionally the ballistic ring can 
be regarded as a network: Each bond 
corresponds to an open mode.
Let us consider the simplest model 
where the scattering is the same for an incident 
particle that comes from the left or from 
the right:
\be{0}
\bm{g} = \left( \amatrix{
[\bm{g}^R]_{a,b}  &  [\bm{g}^T]_{a,b} \cr 
[\bm{g}^T]_{a,b}  & [\bm{g}^R]_{a,b}  }
\right)
\ee
where $\bm{g}^R$ is the reflection matrix
and $\bm{g}^T$ is the transmission matrix. 
The simplest model that one can imagine is with  
\be{9}
\ [\bm{g}^R]_{a,b} &=& \epsilon^2 
\\ \label{e10} 
\ [\bm{g}^T]_{a,b} &=& (1-\mathcal{M}\epsilon^2) \delta_{a,b}
\ee
such that the total transmission is 
\be{0}
g_T = 1-\mathcal{M}\epsilon^2
\ee
Such ``classical" transitions probability matrix 
can arise if we take the $S$ matrix as 
\be{12}
\bm{S}_{\tbox{D}} = \left( \amatrix{
\epsilon\exp\left(i\,2\pi\,\frac{a\, b}{\mathcal{M}}\right) 
& \sqrt{1-\mathcal{M}\epsilon^2} \delta_{a,b} \cr 
\sqrt{1-\mathcal{M}\epsilon^2} \delta_{a,b} 
& -\epsilon\exp\left(-i\,2\pi\,\frac{a\,b}{\mathcal{M}}\right)} 
\right)
\ee
There a lot of simplifications that were 
involved in construction this $\bm{S}$ matrix. 
\begin{itemize}
\item The forward scattering is to the same mode only 
\item The back scattering is ``isotropic"
\item The scattering is energy independent  
\item The scattering phases are not random 
\end{itemize}
One can wonder whether this $\bm{S}$ matrix still qualifies 
as `generic', or maybe the model is over-simplistic. 
In order to illuminate this point let us look 
at the Sinai billiard models of Fig.~2. 
These models are fully qualified as ``quantum chaos" systems.   
One observes that the specific~$\bm{g}$ matrix
of Eq.(\ref{e9}-\ref{e10}) is inspired by that of Fig.~2a. 
In this billiard an incident particle 
is equally likely to be scattered to any mode   
in the backward direction, but the forward scattering 
is only to the same mode (same angle).    
As for the phases: we already have explained 
that our interest is not in disordered ring, 
but rather in chaotic one. Therefore to have 
random phases in the $\bm{S}$ matrix is not an essential 
feature of the model. 
The phases are effectively randomized 
simply because the wavenumber $k_a$ is 
different in each mode. Optionally, it is more 
convenient to assume that all the $k_a$ 
are equal, and instead to have bonds   
of different lengths $L_a$. 
This provides the required phase randomization.

{\bf Generality:}
Though the arguments above strongly suggest 
that the simplified network model is generic,  
we were careful to verify \cite{etzioni} that 
indeed all the results that we find are 
also applicable in the case of the Sinai-type 
system of Fig.~2b where the scatterer is a semi-disc.
We further discuss the {\em robustness} of the results 
in the concluding section. 
Sinai-type billiards are recognized as 
generic chaotic systems. The reason for 
preferring the network model as the leading 
example in the present paper is both 
pedagogical and practical: The mathematics 
is much simpler, and the quality of the numerics 
is much better.

\section{The classical Kubo formula and Drude}

The Fluctuation-Dissipation version of the Kubo formula 
expresses the conductance~$G$ as an integral over the 
current-current correlation function: 
\be{13}
G =  \varrho_{\tbox{F}} \times \frac{1}{2}\int_{-\infty}^{\infty} \langle {\cal I}(t) {\cal I}(0) \rangle dt
\ee
The density of states at the Fermi energy is 
\be{14}
\varrho_{\tbox{F}} 
= \mbox{GeometricFactor} 
\times \mathcal{M} \frac{L}{\pi\hbar v_{\tbox{F}}} 
\ee
where $v_{\tbox{F}}$ is the Fermi velocity.
The geometricfactor depends on the dimensionality $d=1,2,3$. 
The current observable is defined as follows: 
\be{15}
\mathcal{I} = e\hat{v}\delta(\hat{x}-x_0)
\ee
where $\hat{v}$ and $\hat{x}$ are the velocity 
and the position observables respectively. 
In the quantum case a symmetrization of 
this expression is required, so as to obtain an 
hermitian operator. The section through which 
the current is measured 
is arbitrary and we simply take $x_0=+0$, 
namely just to the right of the scattering region.  
Optionally, if there were not the ``black box" region 
of the~$\bm{S}$ matrix, one could average 
over $x_0$, leading to ${\mathcal{I}=(e/L)v}$.

The term ``classical" Kubo formula
implies in this context that the current-current correlation 
function is evaluated classically, ignoring quantum interference. 
In the case of hard chaos system this correlation 
function decays exponentially. The Drude expression  
is the simplest classical approximation:
\be{0}
\langle {\cal I}(t) {\cal I}(0) \rangle = \frac{1}{d}
\left(\frac{e}{L}v_{\tbox{F}}\right)^2 \exp\left[-2\left(\frac{v_{\tbox{F}}}{\ell}\right)|t| \right]
\ee
Substitution into the Kubo formula leads to 
\be{0}
G_{\tbox{Drude}} = \frac{e^2}{2\pi\hbar} \mathcal{M} \frac{\ell}{L}
\ee
where we have dropped the~$d$ dependent prefactor 
which equals~1 for networks ($d=1$). 
In the case of a ballistic ring with 
a restricted scattering region (as in Fig.~1d) 
it is more convenient to characterize the device 
by its total transmission $0<g_T<1$ instead of 
the mean free path. The envelope of 
the current-current correlation function is 
\be{0}
|2g_T-1|^{\tbox{\#rounds}} \ \ \Leftrightarrow \ \ 
\exp\left[-2\left(\frac{v_{\tbox{F}}}{\ell}\right)|t| \right] 
\ee
With the identification of \#rounds as $t/(L/v_{\tbox{F}})$ 
we deduce that for $g_T \sim 1$ the mean free path 
is ${\ell \approx L/(1-g_T)}$. 
A more detailed analysis \cite{kbf} leads to the result   
\be{0}
G_{\tbox{Drude}} &=& \frac{e^2}{2\pi\hbar}
\sum_{a,b} \left[2\bm{g}^T/(1-\bm{g}^T+\bm{g}^R)\right]_{a,b}
\ee
If the device were opened as in Fig.~1e 
we could ignore the multiple rounds. 
In such case we would obtain 
\be{0}
G_{\tbox{Landauer}} &=& \frac{e^2}{2\pi\hbar}
 \sum_{a,b} [\bm{g}^T]_{a,b}
\ee
Both results, the Drude result for the closed 
device and the Landauer result for the 
corresponding open geometry, are ``classical" 
in the sense of Boltzmann, which means that 
they depend only on~$\bm{g}$.

Let us see what do we get for $G_{\tbox{Landauer}}$ 
and for $G_{\tbox{Drude}}$ in the case of the prototype 
system that we have defined in section~2.
The calculation of $G_{\tbox{Landauer}}$ is trivial 
and leads to Eq.(\ref{e3}). 
The calculation of $G_{\tbox{Drude}}$ 
is more complicated since it involves matrix inversion. 
Still~$\bm{g}$ is sufficiently simple to allow 
a straightforward calculation that leads to Eq.(\ref{e4}). 
The rest of this section is devoted to the details 
of this calculation.

We write $\bm{g}_T = \tau^2\bm{1}$ 
and $\bm{g}_R = \epsilon^2 \bm{\Upsilon}$, 
where $\tau$ and $\epsilon$ are defined 
via ${g_T = \tau^2 = 1-\mathcal{M}\epsilon^2}$, 
and where we have introduced the 
following $\mathcal{M}\times\mathcal{M}$ matrices 
\be{0}
\bm{1}
=\left(\begin{array}{cccc}1&0&0&\cdots\\0&1&0&\cdots\\0&0&1&\cdots\\\cdots\\\cdots\end{array}\right)
\,\,\,
\mbox{and}
\,\,\,
\bm{\Upsilon} 
= \left(\begin{array}{cccc}1&1&1&\cdots\\1&1&1&\cdots\\1&1&1&\cdots\\\cdots\\\cdots\end{array}\right)
\ee
Note that the two matrices commute. 
Using these notations we get 
\be{0}
&&
\frac{2\bm{g}_T}{\bm{1}-\bm{g}_T+\bm{g}_R}
=
\frac{2\tau^2}{1-\tau^2}
\frac{1}{\bm{1}+\frac{\epsilon^2}{1-\tau^2}\bm{\Upsilon}}
\\
&&
=
\frac{2\tau^2}{1-\tau^2}
\frac{1}{\bm{1}+\bm{c}\bm{c}^t}
=
\frac{2\tau^2}{1-\tau^2}\,\bm{1}
-\frac{\epsilon^2\tau^2}{(1-\tau^2)^2}
\bm{\Upsilon}
\ee
where we have defined the normalized column vector 
\be{0}
\bm{c}_a=\frac{\epsilon}{\sqrt{1-\tau^2}}
\ \ \ \ \ \ a=1,2,\cdots \mathcal{M} 
\ee
and we have used the identity 
\be{0}
\frac{1}{\bm{1}+\bm{c}\bm{c}^t}
=\bm{1} - \frac{1}{1+\bm{c}^t\bm{c}} \bm{c}\bm{c}^t 
= \bm{1} - \frac{1}{2} \bm{c}\bm{c}^t
\ee
Observing that $\sum_{ab} \bm{1}_{ab} = \mathcal{M}$ 
and $\sum_{ab} \bm{\Upsilon}_{ab} = \mathcal{M}^2$  
we get the desired result Eq.(\ref{e4}).

\section{The quantum Kubo formula and beyond}

Our objective is to find the conductance 
of the closed ring in circumstances such 
that the motion inside the ring is coherent 
(quantum interference within the bonds is not ignored).
The calculation is done using the quantum 
version of Eq.(\ref{e13}) which involves the matrix 
elements $\mathcal{I}_{nm}$ of the current 
operator: 
\be{26}
G = \pi\hbar \,\, \varrho_{\tbox{F}}^2 \times 
\langle\langle |\mathcal{I}_{nm}|^2 \rangle\rangle
\ee
This equation would be the traditional Kubo 
formula if ${\langle\langle...\rangle\rangle}$ 
stood for a simple algebraic average 
over near diagonal matrix elements 
at the energy range of interest. 
By near diagonal elements 
we mean $|E_n-E_m|\lesssim\Gamma$, 
where $\Gamma$ is level broadening parameter.  
The levels of the system are effectively ``broadened" 
due to the non-adiabaticity of the driving~\cite{pmc} 
or due to the interaction 
with the noisy environment~\cite{IS}.
In what follows we assume 
\be{27}
\Delta \ \ \ll \ \ \Gamma \ \ \ll \ \ \Delta_b 
\ee
where $\Delta=1/\varrho_{\tbox{F}}$ is the mean level spacing, 
and $\Delta_b= \pi\hbar v_{\tbox{F}} /L$ is the Thouless energy.
(Note that $\Delta_b/\Delta = \mathcal{M}$). 
Contrary to the naive expectation it has been 
argued in \cite{kbr} that depending on the physical 
circumstances the definition of ${\langle\langle...\rangle\rangle}$ 
may involve a more complicated {\em coarse graining} 
procedure. Consequently the result for~$G$ 
may differ by orders of magnitude 
from the traditional Kubo-Drude result.
We shall discuss this key observation in later sections.

For a network system 
${\varrho_{\tbox{F}} = \mathcal{M} L/(\pi\hbar v_{\tbox{F}})}$.
Furthermore it is convenient to 
define a scaled matrix $I_{nm}$ via the relation 
\be{28}
\mathcal{I}_{nm}=-i(ev_{\tbox{F}}/L)I_{nm} 
\ee
so as to deal with real dimensionless quantities.
Thus we re-write Eq.(\ref{e26}) as: 
\be{29}
G = \frac{e^2}{2\pi\hbar} \ 2\mathcal{M}^2 \ \mathsf{g}
\ee
where $\mathsf{g} \equiv \langle\langle |I_{nm}|^2 \rangle\rangle$. 
In later sections we shall discuss the recipe 
for the~$\mathsf{g}$ calculation. 
It is important to realize (see next section) 
that ${\mathsf{g}<1}$.  
This implies a quantum mechanical bound on~$G$.

\section{The quantum bound on $G$}

We write the channel wavefunctions 
as $\Psi_a(x)=A_a\sin(kx+\varphi)$, 
where $a$ labels the modes.
In our simplified network model 
the modes are re-interpreted as bonds, 
and we assume that the 
wavevector~$k=(2\mass E)^{1/2}$ 
is the same for all bonds. 
For the matrix elements of $\mathcal{I}$ 
we have the expression 
\be{0}
{\cal I}_{nm} \approx  -i e v_{\tbox{F}} \sum_a
\frac{1}{2} \mathsf{A}^{(n)}_a\mathsf{A}^{(m)}_a 
\sin(\varphi^{(n)}_a-\varphi^{(m)}_a)
\ee
were the approximation takes into account 
that our interest is in the couplings 
between levels with ${k_n \approx k_m \approx k_{\tbox{F}}}$. 
From this expression we deduce that 
the scaled matrix elements of Eq.(\ref{e28}) 
are bounded as follows: 
\be{0}
I_{nm}  
\ \ < \ \ 
\sum_a
\frac{L}{2} \mathsf{A}^{(n)}_a \mathsf{A}^{(m)}_a 
\ \ < \ \ 1
\ee
Therefore, irrespective of the details 
of the averaging or coarse graining procedure, 
it is clear that $\bm{g}<1$ as stated at the 
end of the previous section. Consequently  
\be{32}
G\Big|_{\tbox{maximal}} \ \ = \frac{e^2}{2\pi\hbar} \mathcal{M}^2
\ee
In the last expression we have omitted a factor of~$2$.
This is not a typo. We shall explain this point in section~12.

\section{The ergodic result for $G$}

The simplest hypothesis is that all the 
wavefunctions are ergodic random waves.
This is in the spirit of Mott's 
derivation \cite{mott1,mott2}, 
where it has been demonstrated that 
a random wave assumption recovers 
(via Eq.(\ref{e26})) the Drude result.
If indeed the wavefunctions were spread 
equally over all the bonds, it would 
imply ${|\mathsf{A}_a|^2 \sim 2/(\mathcal{M}L)}$. 
If this were true we would get 
\be{0}
|I_{nm}|^2 \ \ = \ \   
\left| \frac{1}{\mathcal{M}}\sum_a  \sin(\varphi^{(n)}-\varphi^{(m)}) \right|^2  
\ \ \approx \ \  \frac{1}{2\mathcal{M}}
\ee
This would imply that the conductance of a ballistic (chaotic) ring is  
\be{34}
G\Big|_{\tbox{ergodic}} \ \ \ = \frac{e^2}{2\pi\hbar} \mathcal{M} 
\ee
We would like to argue that this result 
is wrong. Moreover, it must be wrong.
The result is wrong because the eigenfunction 
of a ballistic ring are not ergodic.  
This we discuss in section~10. 
Furthermore, the result {\em must} be wrong 
because it violates quantum-classical correspondence, 
which we discuss in the next section.

\section{The quantum conductance and Drude}

In this section we define the distinction 
between mesoscopic and spectroscopic conductance 
and further discuss the latter. In the next 
section we elaborate on the calculation procedure of both.
We would like to clarify in advance that the 
spectroscopic conductance is the outcome 
of the traditional Kubo calculation. 
Moreover, it is only the spectroscopic conductance  
which obeys quantum-classical correspondence 
considerations.

{\bf Mesoscopic conductance:} 
If the environmentally induced relaxation 
can be neglected, the rate of energy 
absorption depends on having 
{\em connected} sequences of transitions 
between levels~\cite{kbr}.  
In the next section we explain the proper 
procedure for calculating the conductance 
in such circumstances.  The result 
that comes out from such calculation is what 
we call the mesoscopic conductance  $G_{\tbox{meso}}$.

{\bf Spectroscopic conductance:} 
Within the framework of linear response theory, 
it is assumed that the EMF-induced 
transitions are very slow  
compared with the environmentally 
induced relaxation. 
Then one can argue that Eq.(\ref{e26}) 
is valid with $\langle\langle |\mathcal{I}_{nm}|^2 \rangle\rangle$ 
interpreted as an {\em algebraic average} 
over the matrix elements.  This is what 
we called in section~4 
the {\em traditional} Kubo formula.     
Optionally, if applicable, one may perform 
an algebraic average over realizations of 
disorder. The latter is a very common procedure 
in diagrammatic calculations.   
The outcome of the (traditional) calculation  
is what we call the spectroscopic conductance  $G_{\tbox{spec}}$.  
For further discussion of the conditions that 
justify a ``spectroscopic" calculation see Ref.~\cite{kbr}.

The  spectroscopic conductance  
is not very sensitive to $\Gamma$. 
In fact the $\Gamma$~dependence of the result  
is nothing else but the weak localization 
correction~\cite{kamenev}. 
It scales like $\Delta/\Gamma$ 
for diffusive rings, where $\Delta$
is the mean level spacing (note Eq.(\ref{e27})).

Disregarding weak localization corrections 
it can be argued \cite{kbr} that the obtained 
result for $G_{\tbox{spec}}$ is $G_{\tbox{Drude}}$
provided some reasonable quantum-to-classical 
correspondence conditions (see below) are satisfied.  
It follows that the ergodic hypothesis of the previous 
section cannot be correct, because $G_{\tbox{Drude}}$
is definitely not bounded by the number 
of open modes - it can be much larger.

The necessary condition for quantum-classical 
correspondence can be deducted by taking into 
account the quantum bound of section~5. 
As~$g_T$ becomes closer to~$1$, 
the Drude expression diverges.  
Quantum-to-classical correspondence 
is feasible provided the quantum 
bound is not exceeded:
\be{0}
\frac{1}{1-g_T} \ \ \ll \ \ \mathcal{M}
\ee
This can be re-phrased as 
\be{0}
\frac{\ell}{L} <  \mathcal{M} 
\ee
or as 
\be{0}
t_{cl} \ \ \ll \ \ t_H
\ee
where ${t_{cl} = \ell/v_{\tbox{F}}}$
is the ballistic time, 
and  $t_H=\mathcal{M} \times (L/v_{\tbox{F}})$
is the Heisenberg time
(the time to resolve the quantized energy levels).
In order to establish quantum-to-classical 
correspondence in a constructive manner one 
should express the Kubo formula using 
Green functions, leading to a double summation 
over paths. Then one should argue that energy averaging 
justify the use of the diagonal approximations. 
The procedure is the same as in~\cite{pmt}. 
Needless to say that Eq.(\ref{e34}) is not consistent 
with this argued correspondence and therefore 
a-priori {\em must} be wrong.

\section{The calculation of $G$}

The SLRT recipe for the calculation of~$G$ 
is implied by the following statements:
\begin{itemize}
\item[{[A]}] 
The transitions rates between levels ($w_{nm}$) are given by the FGR.
\item[{[B]}] 
The diffusion in energy ($D$) is given by a resistor network calculation. 
\item[{[C]}] 
The diffusion-dissipation relation 
at low temperatures is $\dot{\mathcal{W}}=\varrho_{\tbox{F}}D$.
\item[{[D]}] 
The expression for the conductance~$G$ is identified via Eq.(\ref{e1}).
\end{itemize} 
In this section we give all the relevant details for using this recipe.
The Hamiltonian of the ring in the adiabatic basis is  
$\mathcal{H} \mapsto E_n\delta_{nm} + W_{nm}$ 
where $W_{nm} = i\dot{\Phi}\hbar\mathcal{I}_{nm}/(E_n{-}E_m)$, 
and $-\dot{\Phi}$ is the EMF. The FGR transition rate 
between level~$n$ and level~$m$ is 
\be{38}
w_{nm} \ \ = \ \ \frac{2\pi}{\hbar} \delta(E_n-E_m) |W_{nm}|^2
\ee
Since we are dealing with a closed system 
one should take explicitly into account 
the broadening of the delta function:
\be{39}
\delta(E_n{-}E_m) \ \ \longmapsto \ \ 
\frac{1}{\Gamma} F\left(\frac{E_n-E_m}{\Gamma}\right)
\ee
The normalized kernel $F()$ reflects either 
the power spectrum or the non-adiabaticity 
of the driving. For the purpose of numerical 
demonstration we assume $F(r)=\exp(-2|r|)$ 
as in~\cite{slr}.  
The level broadening~$\Gamma$ is identical 
with $\Gamma$ of Ref.\cite{kbr,pmc} 
and with $\hbar\omega_0$ of Ref.\cite{slr}. 
As in the conventional derivation of 
linear response theory \cite{kamenev}, 
also here we regard $\Gamma$ as a free 
parameter in the theory.  
It is convenient to use dimensionless 
quantities, so we re-write Eq.(\ref{e38}) as:   
\be{40}
w_{nm} \ \ = \ \ 
\varrho_{\tbox{F}} 
\frac{e^2}{\pi\hbar} 
\mathcal{M}^2 \mathsf{g}_{nm}
\dot{\Phi}^2
\ee
where the dimensionless transition rates are 
\be{41}
\mathsf{g}_{nm} \ \ = \ \ 
2\varrho_{\tbox{F}}^{-3}
\frac{|I_{nm}|^2}{(E_n-E_m)^2} \   
\frac{1}{\Gamma} F\left(\frac{E_n-E_m}{\Gamma}\right)
\ee
In practice we could make the 
approximation $(E_n{-}E_m)/\Delta \approx (n{-}m)$,  
that underestimates exceptionally large couplings
between almost degenerated levels.
Such an approximation would not be reflected  
in the $G_{\tbox{meso}}$ calculation (see below), 
because the latter is determined by the bottlenecks.

The FGR transitions between levels  
lead to diffusion in energy space. 
We would like to calculate 
the {\em coarse grained diffusion coefficient}~$D$
without assuming that all the $|I_{nm}|^2$ 
are comparable. For this purpose it is useful 
to exploit the following resistor network analogy~\cite{slr}: 
\be{0}
w_{nm}^{-1} \ \ \Longleftrightarrow \ \ \mbox{resistor between node $n$ and node $m$} \\
D^{-1}  \ \ \Longleftrightarrow \ \ \mbox{resistivity of the network}
\ee
In dimensionless units~$w_{nm}$ is denoted 
as $\mathsf{g}_{nm}$ as defined via Eq.(\ref{e40}). 
In dimensionless units~$D$ 
is denoted as $\mathsf{g}$ and it is defined 
via the following equation:   
\be{44}
D \ \ = \ \ 
\varrho_{\tbox{F}}^{-1} \frac{e^2}{\pi\hbar} 
\mathcal{M}^2 \mathsf{g}
\dot{\Phi}^2
\ee
The extra $\varrho_{\tbox{F}}^{-2}$ factor compared 
with Eq.(\ref{e40}) arise because the 
resistivity $D^{-1}$ is calculated 
per unit ``energy length" while the scaled 
resistivity $\mathsf{g}^{-1}$ is per unit site.

A standard numerical procedure 
is used for extracting $\mathsf{g}$  
for a given resistor network $\mathsf{g}_{nm}$. 
The steps are as follows: 
{\bf \ (i)} Cut an $N$~site segment out of the 
network (Fig.~3).
{\bf \ (ii)} Define a vector ${\bm{J}_n (n=1..N)}$ 
whose elements are all zero except the first 
and the last that equal ${\bm{J}_1=+J}$ and ${\bm{J}_N=-J}$.
{\bf \ (iii)} Solve the matrix equation 
\be{0}
\bm{J}_n = \sum_{m} \mathsf{g}_{nm} (\bm{V}_n-\bm{V}_m)
\ee
This equation should be solved for $\bm{V}_n$. 
In practice it is easier to write this equation 
as $\bm{J}=-\tilde{\mathsf{g}}\bm{V}$
where ${\tilde{\mathsf{g}}_{nm} = \mathsf{g}_{nm} 
-\delta_{nm} \sum_{m'} \mathsf{g}_{nm'}}$.
The difference $\bm{V}_1-\bm{V}_N$ is obviously proportional 
to the injected $J$. 
{\bf \ (iv)} Find the overall resistance of
the truncated network ${\mathsf{g}_N = J/(\bm{V}_N-\bm{V}_1)}$.
And finally: {\bf \ (v)} Define the resistivity 
as $\mathsf{g}^{-1}=\mathsf{g}_N^{-1}/N$. 
For a locally homogeneous network it has been 
argued in Ref.\cite{kbr} that~$\mathsf{g}$ 
can be obtained via an harmonic average: 
\be{46}
\mathsf{g} \Big|_{\tbox{meso}} \approx 
\left[\frac{1}{N}\sum_{n}^{N} 
\left[\frac{1}{2}\sum_m (m-n)^2 \mathsf{g}_{nm}\right]^{-1}
\right]^{-1}
\ee
The internal sum reflects addition of resistors in parallel, 
while the harmonic average reflects addition 
of resistors in series. This should be contrasted 
with the algebraic average which is used in order 
to calculate the spectroscopic result:
\be{47}
\mathsf{g} \Big|_{\tbox{spec}} = 
\left[\frac{1}{N}\sum_{n}^{N} 
\left[\frac{1}{2}\sum_m (m-n)^2 \mathsf{g}_{nm}\right]
\right]
\ee
It is a simple exercise to verify  
that if all the matrix elements are the same, 
say  $|I_{nm}|^2 = \sigma^2$, 
then ${\mathsf{g}_{\tbox{meso}} = \mathsf{g}_{\tbox{spec}} = \sigma^2}$. 
But if the matrix is structured or sparse 
then ${\mathsf{g}_{\tbox{meso}}}$ is 
much smaller compared with ${\mathsf{g}_{\tbox{spec}}}$. 
Schematically we write in both cases 
\be{0}
\mathsf{g}=\langle\langle |I_{nm}|^2 \rangle\rangle 
\ee
It should be clear that in both cases 
(spectroscopic, mesoscopic) the ``averaging" 
requires the specification of the smoothing scale~$\Gamma$
as implied by Eq.(\ref{e41}). It is also clear 
that the mesoscopic result is much more sensitive 
to the value of $\Gamma$. Unlike the 
spectroscopic result where the dependence 
on~$\Gamma$ is merely a ``weak localization correction" \cite{kamenev}, 
in the case of the mesoscopic result the dependence  
on~$\Gamma$ is a leading order effect.

The diffusion-dissipation relation states 
that the rate of energy absorption 
is ${\dot{\mathcal{W}} = \varrho_{\tbox{F}} D}$. 
Then it is implied by Eq.(\ref{e1}) that the 
conductance~$G$ is given by Eq.(\ref{e29}). 
The procedure above 
can be summarized by saying that $\mathsf{g}$ 
can be calculated from $|I_{nm}|^2$  via
an appropriate ``averaging procedure". 
The appropriate  averaging procedure is 
{\em algebraic} (Eq.(\ref{e47})) in the case 
of the spectroscopic conductance. 
The appropriate  averaging procedure 
is {\em harmonic-type} (as discussed above) 
in the case of mesoscopic conductance.

\section{The eigenstates of the network model}

The network model that we have presented in section~2 
is defined in terms of the scattering 
matrix $\bm{S}_{\tbox{D}}$, 
and the free propagation matrix $\bm{S}_{\tbox{W}}$,
\be{0}
\bm{S}_{\tbox{D}} 
&=& \left( \amatrix{
\epsilon\exp\left(i\,2\pi\,\frac{a\, b}{\mathcal{M}}\right) 
& \sqrt{1-\mathcal{M}\epsilon^2} \delta_{a,b} \cr 
\sqrt{1-\mathcal{M}\epsilon^2} \delta_{a,b} 
& -\epsilon\exp\left(-i\,2\pi\,\frac{a\,b}{\mathcal{M}}\right)} 
\right)
\\
\bm{S}_{\tbox{W}} 
&=& \left( 
\begin{array}{cc} 0 & e^{ikL_a}\,\delta_{ab}\\ 
e^{ikL_a}\,\delta_{ab} & 0 \\ \end{array} 
\right)
\ee
The wavefunction can be written as
\be{0}
|\psi\rangle 
\ \ \longmapsto \ \ \sum_{a=1}^{\mathcal{M}} \left( 
A_{La}  \eexp{i\,k\,(x-L_a)} 
+A_{Ra} \eexp{-i\,k\,x}
\right)
\,\otimes|a\rangle.
\ee
The set of amplitudes $A_L$ and $A_R$ that can be arranged 
as a column vector of length~$2\mathcal{M}$.      
The linear equation for the eigenstates is  
\be{0}
\left( \begin{array}{cc} A_L \\ A_R  \end{array} \right)
=\bm{S}_{\tbox{W}} \, \bm{S}_{\tbox{D}} \,
\left( \begin{array}{cc} A_L \\ A_R  \end{array} \right) 
\label{Eqdot}
\ee
and the associated secular equation for the eigenvalues is 
\be{0}
\det[ \, \bm{S}_{\tbox{W}} \, \bm{S}_{\tbox{D}} - 1 \, ] \ \ = \ \ 0
\ee

In the absence of driving we have 
time reversal symmetry,  
and the unperturbed eigenfunctions can be 
chosen as real (see appendix~A):
\be{0}
|\psi\rangle 
\ \ \longmapsto \ \ 
\sum_{a=1}^{\mathcal{M}}
A_a \sin(kx+\varphi_a) 
\,\otimes|a\rangle.
\ee
The wavefunction is normalized as
\be{0}
\sum_{a=1}^{\mathcal{M}} \int_0^{L_a} A_a^2 \sin^2(kx+\varphi_a)\,dx = 1 
\ee 
which implies
\be{0}
\sum_{a=1}^{\mathcal{M}} \frac{L_a}{2} A_a^2 \approx 1 
\ee 
For a given $g_T$ we can find  
numerically the eigenvalues 
and the eigenstates, thus 
obtaining a table    
\be{0}
(k_n, \varphi_a^{(n)}, A_a^{(n)})
\ \ \ \ \ \ \ \ n=\mbox{level index}
\ee
For the numerical study we have chosen a network 
system consisting of $\mathcal{M}=50$ bonds.
The length of each bond is randomly selected  
in the range $L_a=1\pm 0.1$. 
We select the eigenvalues with $k_n \sim 2000$. 
The numerical results over the 
whole range of $g_T$ values are 
presented in Figs.~4-7. In the following 
sections we discuss and analyze 
these results.

It is of course possible to determine   
analytically what are the eigenvalues 
and the eigenstates in the $g_T \rightarrow 1$ limit. 
The combined scattering matrix is
\be{0}
\bm{S}_{\tbox{W}}\,\bm{S}_{\tbox{D}}
=\left(\begin{array}{cccc}
\tau\,\eexp{ikL_a}\delta_{a,b} 
&  &  & -\epsilon\,\eexp{i(kL_a-\frac{2\pi}{\mathcal{M}}a\times b)} \\
\epsilon\,\eexp{i(kL_a\,+\frac{2\pi}{\mathcal{M}} a\times b)} 
&  &  & \tau\,\eexp{ikL_a}\delta_{a,b}
\end{array}\right)
\ee
where we use the notation $\tau=(1-\mathcal{M}\epsilon^2)^{1/2}$.
For $g_T=1$ this matrix becomes diagonal.  
Then it has $\mathcal{M}$ distinct eigenvalues, 
each doubly degenerate. 
We are interested in the non-degenerate case 
in the limit $\epsilon \rightarrow 0$. 
The eigenstates are still localized each 
in a single $a$~bond, 
but the  degeneracy is lifted.  
Within the framework of degenerate perturbation 
theory we have to diagonalize the~${2 \times 2}$ matrix 
\be{0}
\left(\begin{array}{cc}
\tau\,\eexp{ikL_a} 
& -\epsilon\,\eexp{i(kL_a\,-\frac{2\pi\,a^2}{\mathcal{M}})}\\
\epsilon\,\eexp{i(kL_a\,+\frac{2\pi\,a^2}{\mathcal{M}})} 
& \tau\,\eexp{ikL_a}
\end{array}\right)
\ee
whose eigenvalue are determined 
by the associated secular equation 
\be{0}
(\epsilon^2 + \tau^2)\eexp{2ikL_a} - 2\tau \eexp{ikL_a} + 1 = 0
\ee
Hence we get the following approximations 
\be{0}
k_n &\approx& 
\left(
2\pi \times\mbox{\small integer} 
\pm \frac{1}{\sqrt{\mathcal{M}}} \ \epsilon
\right)
\frac{1}{L_a}
\\ \label{e62}
\varphi_a^{(n)} & \approx &
-\frac{a^2}{\mathcal{M}}\pi
-\frac{1}{2}k_nL_a
+\left\{ \begin{array}{cc}\pi/4\\3\pi/4\end{array} \right.
\ee
We have verified that the numerical results of Fig.~4 and Fig.~7
agree with these estimates. We note that 
for hard wall scatterer each $\varphi_a$ 
would become either $0$ or $\pi/2$ 
in the $g_T \rightarrow 1$ limit.

\section{The non-ergodicity of the eigenfunctions}

In Fig.~5 we display images of the column vectors $A_a^{(n)}$ 
for two representative values of $g_T$ 
so as to illustrate the crossover from 
localized to ergodic wavefunctions. 
Each eigen-function can be characterized  
by its participation ratio:
\be{0}
\mbox{PR} = 
\left[
\sum_a  \left(\frac{L_a}{2} A_a^2 \right)^2
\right]^{-1}
\ee
This constitutes a measure for the 
ergodicity of the eigen-functions. 
By this definition 
\be{0}
\mbox{PR} \approx 
\left\{ \amatrix{ 
1 & \mbox{for a single bond localized state} \cr 
\mathcal{M} & \mbox{for a uniformly distributed state}  
} \right.
\ee
We distinguish between 3~regimes depending on 
the value of the total transmission~$g_T$,  
\begin{itemize}
\item The trivial ballistic regime $(1-g_T) \ll 1/\mathcal{M}$ for which $\mbox{PR} \sim 1$
\item The non-trivial ballistic regime $1/\mathcal{M}\ll (1-g_T)\ll 1$. 
\item The non-ballistic regime where $g_T$ is not close to $1$ and $\mbox{PR} \sim \mathcal{M}$
\end{itemize} 
In the trivial ballistic regime 
the eigenstates are like those of 
a reflection-less ring with uncoupled modes,
hence $\mbox{PR} \sim 1$. Once $(1-g_T)$ becomes 
larger compared with $1/\mathcal{M}$ first order 
perturbation theory breaks down, 
and the mixing of the levels is described 
by a Wigner Lorentzian. The analysis is completely 
analogous to that of the single mode case of Ref.\cite{kbr}, 
leading to $\mbox{PR} \propto (1-g_T) \times \mathcal{M}$.  
For $g_T$ values that are not close to~$1$ 
the eigen-functions become ergodic 
with $\mbox{PR} \sim \mathcal{M}$. 
From RMT we expect \cite{ParticRatio} $\mbox{PR} \sim \mathcal{M}/3$.
A satisfactory global fit, that works well 
within the non-trivial ballistic regime is (Fig.~8):   
\be{65}
\mbox{PR} 
\approx  1 + \frac{1}{3}(1-g_T) \mathcal{M}
\ee
Our interest 
is focused in the {\em non-trivial} ballistic   
regime $1/\mathcal{M} \ll  (1-g_T) \ll 1$, 
where we have strong mixing of 
levels ($\mbox{PR} \gg 1$), but still the 
mean free path $\ell \approx L/(1-g_T)$  
is very large compared with the 
ring's perimeter ($\ell \gg L$).
In view of the discussion in section~6, 
it is important to realize that in this regime 
we do not have ``quantum chaos" ergodicity.   
Rather we have $\mbox{PR} \ll \mathcal{M}$ 
meaning that {\em the wavefunctions occupy only a small 
fraction  of the classically accessible phase space}.

\section{The calculation of matrix elements}

Given a set of eigenstates, it is 
straightforward to calculate the matrix elements 
of the current operator (Figs.~9-12). 
We recall that the 
scaled matrix elements are 
\be{0}
I_{nm} \approx  
\sum_a \frac{L_a}{2} A_a^{(n)} A_a^{(m)} 
\sin(\varphi_a^{(n)} - \varphi_a^{(m)}) 
\ee
with the associated upper bound 
\be{0}
\bar{I}_{nm} \approx  
\sum_a \frac{L_a}{2} A_a^{(n)} A_a^{(m)} 
\ee
For $n=m$ we have $\bar{I}_{nm}=1$ due to normalization, 
and  $I_{nm}=0$ due to time reversal symmetry. 
From now on we are interested in $n \neq m$.
There are several extreme cases 
that allow simple estimates: 
\be{-1}
\hspace*{-2cm}
\bar{I}_{nm} \approx \left\{ 
\begin{array}{cl} 
0  \hspace{3mm} & 
\mbox{for pair of states localized on different bonds} 
\\
1  \hspace{3mm} &
\mbox{for pair of states localized on the same bond}
\\ 
1  \hspace{3mm} &
\mbox{for pair of ergodic states} 
\end{array} \right.
\ee
If we take the phases into account we get
\be{-1}
\hspace*{-2cm}
|I_{nm}|^2 \approx \left\{ 
\begin{array}{cl} 
0  \hspace{3mm} & 
\mbox{for pair of states localized on different bonds} 
\\
1  \hspace{3mm} &
\mbox{for pair of nearly degenerated states on the same bond} 
\\
1/(2\mathcal{M}) \hspace{3mm} &
\mbox{for pair of uncorrelated ergodic states} 
\end{array} \right.
\ee
We have already explained in section~6 
that the ``ergodic" hypothesis 
is wrong in the ballistic case. 
It should be clear that the small PR of the 
eigenfunctions implies sparsity of $I_{nm}$: 
The matrix elements are very small for 
any pair of states that are localized 
on different sets of bonds. 
This observation is demonstrated in Figs.~9-12.
As the reflection $1-g_T$ is increased, 
more and more elements become non-negligible, 
and the matrix becomes less structured and less sparse.

\section{Numerical results for the conductance}

Once we have the matrix elements $|I_{nm}|^2$
we can calculate $G_{\tbox{spec}}$ using  
the algebraic average recipe Eq.(\ref{e47}).  
We can also calculate $G_{\tbox{meso}}$ using 
either the resistor network procedure 
or the harmonic average approximation Eq.(\ref{e46}).
Fig.~13 displays the results for an $\mathcal{M}=50$ 
network model. The conductance goes to zero 
for both ${g_T \rightarrow 0}$ and ${g_T \rightarrow 1}$. 
The dependence on $\Gamma$ is plotted in Fig.~14. 
The rough accuracy of the harmonic average has 
been verified (not displayed).

The dependence of $G$ on the smoothing 
parameter~$\Gamma$ is easily understood 
if we keep in our mind the band profile 
which is illustrated in Fig.~10. 
In order to improve our intuition 
we show in Fig.~12 the average value 
of $|I_{n,n+r}|^2$ for $r=1,2,3,4,5$
as a function of $1-g_T$ for ${r=1,2,3,4,5}$.

It should be clear that the large $r=1$ 
elements originate from the pairs of 
almost degenerate states that were discussed 
in section~9. Their contribution 
to the spectroscopic conductance 
is dominant. The upper bound Eq.(\ref{e32}) 
on $G$ is implied by the upper bound 
on $|I_{n,n+1}|^2$. It was already pointed 
out in section~11 that the maximal  
value $|I_{nm}|=1$ is  attained for the 
nearly degenerate states. The algebraic 
average with the interlacing vanishingly  
small couplings leads to the factor of $1/2$  
that was mentioned after Eq.(\ref{e32}). 
To avoid miss-understanding we emphasize 
that this {\em prefactor} is model specific.

On the other hand, the large $r=1$ couplings  
almost do not affect the mesoscopic 
conductance. This is because they do not form 
connected sequences.  Moreover, as implied 
by our calculation recipe, large value 
of $\Gamma$ cannot help to overcome the  
bottlenecks. In order to get a classical result 
the environment should induce not only 
level broadening (which is like the $1/T_2$ 
rate of pure dephasing in NMR studies), 
but also a relaxation effect
(analogous to the $1/T_1$ rate in NMR).

\newpage
\section{Discussion}

In this paper we have studied the mesoscopic 
conductance of a ballistic ring 
with mean free path $\ell \gg L$.
The specific calculation has been done for 
a network model, but all its main ingredients 
are completely {\em generic}. 
Ballistic rings with $\ell \gg L$ are not 
typical ``quantum chaos" systems. 
Their eigenfunctions {\em are not ergodic over 
the whole accessible phase space}, 
and cannot be regarded as 
an extended ``random wave". Consequently  
the perturbation matrix $\mathcal{I}_{nm}$
is highly structured and sparse, 
and we have to go beyond the Kubo formalism    
in order to calculate the mesoscopic conductance.

{\bf Results vs expectations:}
The ``averaging" over the matrix elements 
of the current operator should be done 
according to the appropriate prescription: 
algebraic scheme for the spectroscopic 
conductance $G_{\tbox{spec}}$, 
and resistor-network scheme for the mesoscopic 
conductance $G_{\tbox{meso}}$. 
The calculation procedure implies that 
\be{0}
G_{\tbox{meso}} < G_{\tbox{spec}} \le  G_{\tbox{Drude}}
\ee
Our original naive belief, before 
we started with the numerical work, 
was that it is feasible to get 
quite large $G_{\tbox{meso}}$, 
possibly of the order of $G_{\tbox{spec}}$. 
To our surprise the numerics has revealed 
that {\em typically} 
\be{0}
G_{\tbox{meso}} <  G_{\tbox{Landauer}}
\ee
We have pushed our numerical verification 
of this statement up to ${\mathcal{M}=450}$ (Fig.~14b).
For an optimal value of $\Gamma$, such 
that $G_{\tbox{meso}}$ is maximal, we still 
have  ${G_{\tbox{meso}} \lesssim  G_{\tbox{Landauer}}}$.
The numerical prefactor in the latter inequality 
appears to be roughly~$3$, but obviously 
we cannot establish that there is a strict limitation.      
Still, as far as order of magnitude estimates 
are concerned, our conjecture is that this 
statement is true {\em in general}. 
We did not find a mathematical argument 
to establish this conjecture, except the 
very simple case of a single mode ballistic 
ring\cite{kbr} where the calculations 
of~$G$ can be done analytically.


{\bf Robustness:}
Our results are not sensitive to the details of the model. 
Disregarding the details, the eigenfunctions 
are doomed to be non-ergodic in mode space 
if $g_T\sim1$. This by itself implies that $I_{nm}$
is sparse and possibly structured. 
Consequently the resistor network picture 
implies that $G_{\tbox{meso}}$ is much smaller 
compared with the naive expectation.   
We have further tested the generality of our 
quantitative statements by analyzing \cite{etzioni} another, 
more realistic model, 
where the ring is modeled as a waveguide with 
a semi-disc scatterer (Fig.~2b). 
The $\bm{S}$ matrix for this model 
is known \cite{blumel,holger}.
In particular we have verified that 
the participation ratio of the eigenstates 
has roughly the expected dependence on~$g_T$. 
Indeed for both the semi-disc model, 
and our simplified network model,  
the participation ratio does not exhibit 
anomalous saturation as typical, say, 
for a ``star graph" \cite{starg}. 
Eventually we have verified that the results 
for the mesoscopic conductance of the 
semi-disc model are similar to those that 
were obtained for the network model.

{\bf Challenges:} 
It is still an open challenge to 
derive an estimate for the mesoscopic 
conductance in terms of~$g_T$.  
It was possible to derive such an expression 
in the single mode case. There we have 
found that ${G_{\tbox{meso}} \propto (1-g_T)^2g_T}$. 
In the general case (${\mathcal{M} > 1}$) 
the calculation is more complicated.  
We suspect that our expression for the participation 
ratio Eq.(\ref{e65}) constitutes an important step towards 
this goal. In any case we were not able to derive 
a reliable closed analytical expression.

{\bf Limitations:} 
It should be emphasized that if 
there is either a very effective 
relaxation or decoherence process, then the 
semi-linear response theory that we have discussed 
do not apply. In the presence of strong environmental 
influence one can justify, depending 
on the {\em circumstances} \cite{kbf}, 
either the use of the traditional 
Kubo-Drude result, or the use of the Landauer result.

\appendix
\section{Implications of time reversal symmetry}

We can decompose the eigenstate equation as follows:
\be{0}
\left( \begin{array}{c} B_{La} \\ B_{Ra} \end{array} \right) 
&=&S_{\tbox{D}} \left( \begin{array}{c} A_{La}\\ A_{Ra} \\ \end{array} \right)
\\
\left( \begin{array}{c} A_{La}\\ A_{Ra} \\ \end{array} \right)
&=&S_{\tbox{W}} \left( \begin{array}{c} B_{La} \\ B_{Ra} \end{array} \right)
\ee   
Above $B_{La}$ and $B_{Ra}$ are the amplitudes 
of the outgoing waves from $x=0$, while $A_{La}$ and $A_{Ra}$ 
are the amplitudes of the ingoing waves.
Conventional time-reversal-symmetry implies that 
both $\psi(x)$ and its complex-conjugate $\psi(x)^*$
satisfy the same Schr\"odinger equation.  
Complex conjugation turns out the incoming wave into 
outgoing one and vice versa, and therefore  
\be{0}
\left( \begin{array}{c} A_{La}^* \\ A_{Ra}^* \end{array} \right) 
&=&S_{\tbox{D}} \left( \begin{array}{c} B_{La}^*\\ B_{Ra}^* \\ \end{array} \right) 
\\
\left( \begin{array}{c} B_{La}^*\\ B_{Ra}^* \\ \end{array} \right)
&=& S_{\tbox{W}} \left( \begin{array}{c} A_{La}^* \\ A_{Ra}^* \end{array} \right) 
\ee
It is not difficult to see that the two sets 
of equations are equivalent provided 
\be{0}
S_{\tbox{D}}^{\tbox{transposed}} &=& S_{\tbox{D}} 
\\
S_{\tbox{W}}^{\tbox{transposed}} &=& S_{\tbox{W}} 
\ee
If we have this (conventional) time reversal symmetry,   
the unperturbed eigenfunctions can be 
chosen as real in position representation:
\be{0}
|\psi\rangle 
=\sum_{a=1}^{\mathcal{M}} 
A_a \sin(kx+\varphi_a) 
\,\otimes|a\rangle.
\ee
where
\be{0}
A_a &=& 2|A_{La}| = 2|A_{Ra}|
\\
\varphi_a &=& 
\frac{1}{2}
\left(\pi+\arg(A_{La}/A_{Ra})-kL_a\right) 
\ee

\clearpage

\ack

YE thanks Itamar Sela for his help to correct 
a subtle numerical error: some data points in 
the original numerical work of Ref.\cite{bls} 
were distorted. 
Much of the motivation for this work came from 
intriguing meetings of DC during 2004-2005 
with Michael Wilkinson, who highlighted 
the open question regarding the 
feasibility to get $G>G_{\tbox{Landauer}}$ 
in the case of a multimode closed ring. 
We also thank Bernhard~Mehlig, Tsampikos~Kottos 
and Holger~Schanz for inspiring discussions. 
The research was supported by the 
Israel Science Foundation (grant No.11/02) 
and by a grant from the DIP, 
the Deutsch-Israelische Projektkooperation.


\end{document}